\documentclass[aps,prb,twocolumn]{revtex4} 
\usepackage{amsmath} 
\usepackage{graphicx} 
\usepackage{color}

\begin{document} 
\title{Magnetoelectric coupling in superconductor-helimagnet heterostructures} 

\author{Kjetil M. D. Hals} 
\affiliation{Niels Bohr International Academy and the Center for Quantum Devices, Niels Bohr Institute, University of Copenhagen, 2100 Copenhagen, Denmark} 
%%%%%%%%%%%%%%%%%%%%%%%%%%%%%%%%%%%%%%%%%%%%%%%%%%%%%%%%%%%%%%%%%%%%%%%%%%%%%%% 
\begin{abstract}
The Ginzburg-Landau free energy of a conventional superconductor coupled to a helimagnet is microscopically derived using functional field integral techniques. We show that the spin texture leads to a Lifshitz invariant in the free energy, which couples the momentum density of the superconducting condensate to the magnetization of the helimagnet. For helimagnets with a conical texture, the Lifshitz invariant yields a spatial modulation of the superconducting phase along the helical wavevector of the magnetic texture. Based on self-consistent numerical calculations, we verify the theoretical formalism by investigating a superconductor that contains a helical Yu-Shiba-Rusinov (YSR) chain. We demonstrate that the texture-induced magnetoelectric coupling produces a strong supercurrent along the YSR chain, which induces a detectable magnetic field.  
\end{abstract}

\maketitle 

%%%%%%%%%%%%%%%%%%%%%%%%%%%%%%%%%%%%%%%%%%%%%%%%%%%%%%%%%%%%%%%%%%%%%%%%%%%%%%% 
\section{Introduction} \label{Sec:Intro}
%%%%%%%%%%%%%%%%%%%%%%%%%%%%%%%%%%%%%%%%%%%%%%%%%%%%%%%%%%%%%%%%%%%%%%%%%%%%%%% 
Heterostructures of conventional superconductors and ferromagnets are currently attracting considerable interest because of their potential use for realizing topological superconductivity.~\cite{Nadj-Perge:PRB2013, Klinovaja:PRL2013, Vazifeh:PRL2013, Simon:PRL2013, Nadj-Perge:Science2014, Pientka:prb2013, Choy:prb2011, Schecter:prb2016, Sato:prb2010, Sau:prl2010, Mao:prb2010, Ojanen:prb2014, Ojanen:prb2014R, Bjornson:prb2013, Nakosai:prb2013, Christensen:arXiv2016,Loss:prb2016} The essential ingredients for achieving a topological phase in these systems are spin-orbit coupling (SOC) combined with magnetism, which lead to an effective p-wave pairing in the superconductor.

Recently, textured ferromagnets have been presented as an alternative approach to achieve p-wave pairing and topological superconductivity even in absence of SOC. In particular, systems consisting of one-dimensional (1D) arrays of magnetic atoms on the surface of a conventional superconductor have been a focus of this discussion,~\cite{Nadj-Perge:PRB2013, Klinovaja:PRL2013, Vazifeh:PRL2013, Simon:PRL2013, Nadj-Perge:Science2014, Pientka:prb2013, Choy:prb2011, Ojanen:prb2014, Ojanen:prb2014R, Schecter:prb2016, Christensen:arXiv2016} but two-dimensional (2D) magnetic layers with skyrmion textures have also been proposed as possible building blocks.~\cite{Nakosai:prb2013, Loss:prb2016} 
The 1D spin arrays are often referred to as Yu-Shiba-Rusinov (YSR) chains.  One reason for the interest in the YSR systems is that the spin chains are predicted to form a spin helix via indirect exchange interactions mediated by the itinerant electrons. In certain parameter regimes, which are determined by, e.g., the chemical potential, the superconducting pair potential, and the coupling strength between the electrons and the spin chain, the spin helix alone drives the superconductor into a topological phase.~\cite{Klinovaja:PRL2013,Vazifeh:PRL2013,Simon:PRL2013,Schecter:prb2016,Christensen:arXiv2016} 
The physical mechanism underlying this self-organized topological phase is the effective SOC produced by the spin texture.  As a simple illustration of this effect, consider the single-particle Hamiltonian $\hat{H}= \hat{\mathbf{p}}^2/2m + \mathbf{h}\cdot \hat{\boldsymbol{\sigma}} $ where $\hat{\mathbf{p}}$ is the momentum operator, $\hat{\boldsymbol{\sigma}}$ is the vector of Pauli matrices, and $\mathbf{h}$ is the helical Zeeman field produced by the helical ordered spins. The effects of the spatially varying spin texture can be absorbed into the momentum operator via an SU(2) gauge transformation $\hat{U}$ that aligns $\mathbf{h}$ along $z$: $\hat{\mathbf{p}} \rightarrow \hat{\mathbf{p}} - \hat{\mathbf{A}}$. Here,  $\hat{A}_k= i\hbar \hat{U}\partial_k \hat{U}^{\dagger}= a_k^{j}\hat{\sigma}_{j}$ is a SU(2)-valued gauge potential, which couples the momentum to the spatial variations of $\mathbf{h}$. Effectively, this SU(2) gauge potential acts as a spatially asymmetric SOC $\sim a_k^{j}\hat{\sigma}_{j} \hat{p}_k$, which combined with the Zeeman splitting can bring the system into a topological state. 

So far, most studies have focused on how this effective SOC affects the topological state of the superconductor. However, there are several indications that the texture also leads other important phenomena in the superconductor. In  Ref.~\onlinecite{Pientka:prb2013}, a tight-binding Bogoliubov-de Gennes (BdG) Hamiltonian was derived for a helical YSR chain. In the case of a conical helix, the hopping amplitudes in the effective Hamiltonian involve complex phase factors, which indicate that the texture leads to spatial modulations of the superconducting phase and supercurrents in the condensate. Magnetic textures also significantly influence the Josephson effect in ferromagnetic Josephson junctions.~\cite{Robinson:science2010,Yokoyama:prb15} Ref.~\onlinecite{Robinson:science2010} carried out measurements on junctions with the conical magnet holmium (Ho) in the magnetic interlayer. They observed peaks in the critical current for thicknesses of the Ho layer corresponding to noninteger spiral wavelengths of the conical texture. These results demonstrate that improved insight into the complex interplay between the magnetization gradients and the condensate is essential for developing a deeper understanding of superconductor-ferromagnet hybrid structures. In the present work, we investigate how the effects of helimagnetic textures can be incorporated into a phenomenological Ginzburg-Landau (GL) description of superconductor-helimagnet heterostructures. 

The GL theory provides a powerful formalism to describe superconductivity.~\cite{Tinkham:book} The starting point of the theory is the formulation of the GL free-energy functional $F$. Near the superconducting transition, $F$ can be expressed as a series expansion in the order-parameter field. For a conventional superconductor, the GL free-energy density $\mathcal{F}$ (related to the free-energy functional via $F[\Psi^{\ast} , \Psi, \mathbf{A} ]= \int {\rm d\mathbf{r}}\mathcal{F} (\mathbf{r})$) takes the form of~\cite{Tinkham:book}
\begin{equation}
\mathcal{F} (\mathbf{r}) =  \frac{1}{4m} (\boldsymbol{\Pi}\Psi)^{\ast}\cdot  \boldsymbol{\Pi}\Psi + a |\Psi|^2 + \frac{b}{2} |\Psi|^4 + \frac{B^2}{ 8\pi} , \label{Eq:Fe}
\end{equation}
where $a$ and $b$ determine the absolute value $|\Psi (\mathbf{r})|$ of the complex order parameter field $\Psi (\mathbf{r})= |\Psi (\mathbf{r})|\exp (i \phi (\mathbf{r}))$, $\boldsymbol{\Pi} = -i \hbar \boldsymbol{\nabla} - (2 e/c) \mathbf{A}$ is the momentum operator, $2m$ ($2e$) is the mass (charge) of a Cooper pair, $c$ is the speed of light, and $\mathbf{A}$ is the magnetic vector potential, which is related to the magnetic induction $\mathbf{B}$ by $\boldsymbol{\nabla}\times\mathbf{A}= \mathbf{B}$. The equilibrium state of the superconductor is given by the GL equations, which are obtained by a variational minimization of $F$ with respect to $\Psi^{\ast}$ and $\mathbf{A}$.

Without the magnetic vector potential, the free-energy density in Eq.~\eqref{Eq:Fe} is invariant under any proper or improper rotation $\boldsymbol{\mathcal{R}}\in {\rm O(3)}$ of the spatial coordinates: $\mathcal{F} (\mathbf{r})= \mathcal{F} (\boldsymbol{\mathcal{R}}\mathbf{r})$. Thus, the free-energy density describes a fully isotropic system. Therefore, Eq.~\eqref{Eq:Fe} does not provide a correct phenomenological description of superconductors with strong SOC and broken spatial inversion symmetry. For such systems, spatially asymmetric terms are allowed in the free energy.~\cite{Mineev:94, Edelstein:96} These terms are often referred to as Lifshitz invariants. In general, the Lifshitz invariant can be phenomenologically written as~\cite{Mineev:94}
\begin{equation}
\mathcal{F}_{me}= -\sum_{ij} \kappa_{ij} h_i  \mathcal{P}_j ,  \label{Eq:Fme}
\end{equation}
where $\mathbf{h}$ is the Zeeman field experienced by the itinerant electrons, and $\boldsymbol{\mathcal{P}}=  \psi^{\ast} \boldsymbol{\Pi} \psi + \psi \boldsymbol{\Pi}^{\ast} \psi^{\ast}$ is the momentum density of the superconducting condensate. Typically, the Zeeman field is produced by an external magnetic field or the exchange field of an adjacent ferromagnetic layer, which is induced by proximity to the superconductor. The tensor $\kappa_{ij}$ is linear in the SOC and is a second-rank axial tensor with a tensorial form, which is determined by the equations $\kappa_{ij}= |\mathcal{R}^{(k)}| \mathcal{R}^{(k)}_{im} \mathcal{R}^{(k)}_{jn}\kappa_{mn}$. Here, $\boldsymbol{\mathcal{R}}^{(k)}\in \mathcal{G}$ ($k= 1, 2, ...$) are the generators of the system's point group $\mathcal{G}$, and $|\boldsymbol{\mathcal{R}}|$ is the determinant. Importantly, $\kappa_{ij}$ vanishes for symmetry groups that contain the inversion operator $\mathcal{R}_{ij}= -\delta_{ij}$ ($\delta_{ij}$: Kronecker delta). 

Because the Lifshitz invariant is linear in the spatial gradients, it favors a helical modulation $\Psi\sim \exp (i \mathbf{Q}\cdot\mathbf{r})$ of the order-parameter field and can lead to the formation of a helical superconducting phase in non-centrosymmetric superconductors subjected to external magnetic fields.~\cite{Mineev:94, Agterberg:physica03, Dimitrova:JETP03} This state is analogous to the helimagnetic order, which is observed in magnetic systems with broken spatial inversion symmetry.~\cite{Dzyaloshinsky:1958, Moriya:1960} In addition, Eq.~\eqref{Eq:Fme} introduces magnetoelectric effects, such as the spin-galvanic effect~\cite{Edelstein:96,Balatsky:prl15} and its reciprocal process.~\cite{Edelstein:prl95} Recently, the effects of the magnetoelectric coupling were theoretically investigated in hybrid systems of coupled superconducting and ferromagnetic layers, where the Lifshitz invariant is predicted to produce several new and exciting phenomena, such as persistent currents generated by ferromagnetism,~\cite{Balatsky:prl15,Malshukov:prb16} composite topological excitations,~\cite{Hals:arxiv16} and supercurrent-induced spin-orbit torques.~\cite{Hals:prb16} 

One intriguing question that remains largely unexplored is how a helical spin texture affects the GL free energy of the superconductor in hybrid superconductor-ferromagnet systems. Because the Zeeman field induced in the superconductor by the spin helix breaks the spatial inversion symmetry of the superconductor and produces an effective SOC, it is natural to ask whether the texture introduces a magnetoelectric coupling term similar to Eq.~\eqref{Eq:Fme} in the GL free-energy functional. 

\begin{figure}[t!] 
\centering 
\includegraphics[scale=1.0]{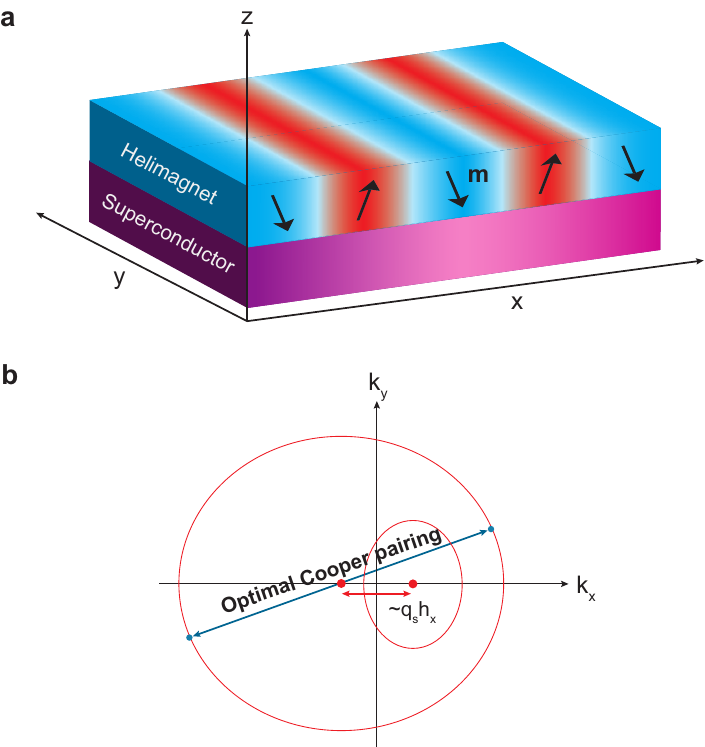}  
\caption{(Color online). (a) A superconductor-helimagnet heterostructure in which the exchange field of the helimagnet is induced by proximity to a thin-film superconductor. 
When the helimagnet contains a conical spin texture (a tilted helix), the texture induces a magnetoelectric coupling between the magnetization $\mathbf{m}$ (tilted arrows) of the magnet and the momentum density of the superconducting condensate. 
This coupling generates a phase gradient (color gradient) in the superconducting layer.
(b) An illustration of the Fermi surface of the Hamiltonian $ \hat{\mathcal{H}}_u$.  In the rotated frame, the electrons experience an effective SOC, which leads to optimal Cooper pairing for pairs of electrons with a finite center-of-mass momentum. 
Phenomenologically, the interaction between the spin texture and the Cooper pairs is captured by a Lifshitz invariant in the GL free energy.  }
\label{Fig1}
\end{figure}

In this paper, we microscopically derive the GL free energy of a heterostructure consisting of a conventional superconductor interfaced with a helimagnet, where
the helimagnet induces a Zeeman field $\mathbf{h}$ in the superconductor via the exchange proximity effect (Fig.~\ref{Fig1}a). Interestingly, we find that the spin texture produces a Lifshitz invariant in the free energy, which couples the helimagnet to the momentum density of the condensate.
For a tilted helix (i.e., a conical texture) with a Bloch structure and a helical wavevector $\mathbf{q}_s= q_s\hat{\mathbf{x}}$, the Lifshitz invariant takes the form of:
\begin{equation}
\mathcal{F}_{me}= -\frac{1}{2}\kappa h_x  \mathcal{P}_x ,  \label{Eq:Fme2}
\end{equation}
where the magnetoelectric coupling constant $\kappa = \hbar q_s / 8\mu_F m$ is governed by the wavevector of the helix, the effective mass $m$ of the itinerant electrons, and the chemical potential $\mu_F$. 
Here, $h_x$ determines the tilting of the helix, which can be tuned by applying a weak magnetic field along $x$. 
To investigate the effects of the texture-induced magnetoelectric coupling, we study a helical YSR chain on a two-dimensional (2D) superconductor and
show that the texture induces a supercurrent and a phase gradient along the chain. 
Importantly, we find that the induced supercurrent generates a magnetic field that can be detected via imaging techniques and, thus, may offer a new and relatively direct method of probing the helical ordering of self-organized topological systems.  

Our findings demonstrate that the Lifshitz invariant \eqref{Eq:Fme2} may play an important role in several different types of hybrid systems of coupled superconducting and ferromagnetic materials.
In helical YSR chains stabilized by electron-induced exchange interactions, $q_s$ can be on the order of the Fermi wavevector $k_F$. For instance, in 1D electron systems (and some 2D systems close to half filling), $q_s = 2k_F$.~\cite{Klinovaja:PRL2013,Vazifeh:PRL2013,Simon:PRL2013,Schecter:prb2016,Christensen:arXiv2016} In this case, the helix produces a strong effective SOC with a SOC length of $\sim 1/2k_F$, which can strongly affect the properties of the YSR systems.  
Thin-film superconductors interfaced with chiral magnets represent another class of heterostructures, in which the texture-induced magnetoelectric coupling is likely crucial for understanding the superconducting properties. Unlike the YSR chains, the spin texture in chiral magnets is stabilized by the Dzyaloshinskii-Moriya interaction (DMI),~\cite{Dzyaloshinsky:1958, Moriya:1960} which is produced by the interfacial or bulk SOC. Consequently, the wavevector $q_s$ becomes linear in the SOC, implying that the coupling parameter $\kappa$ in Eq.~\eqref{Eq:Fme2} can be comparable in magnitude with the conventional magnetoelectric coupling induced by the SOC. Thus, in superconductor/chiral magnet heterostructures, both SOC- and texture-induced Lifshitz invariants must be considered to provide a correct description of the superconductor.  

This paper is organized as follows. In Sec.~\ref{Sec:MicroDerivation}, we microscopically derive a GL theory of superconductor-helimagnet heterostructures. 
In Sec.~\ref{Sec:NumStud}, the effects of the texture-induced magnetoelectric coupling are studied via a numerical and self-consistent calculation of the pair potential and supercurrent  for a helical YSR chain on a superconductor.
The results are summarized in Sec.~\ref{Sec:Summary}.

%%%%%%%%%%%%%%%%%%%%%%%%%%%%%%%%%%%%%%%%%%%%%%%%%%%%%%%%%%%%%%%%%%%%%%%%%%%%%%%%%%%%%%%%%%%%%%%%%%%%%%%%%%%%%%%% 
\section{Microscopic Derivation} \label{Sec:MicroDerivation}
%%%%%%%%%%%%%%%%%%%%%%%%%%%%%%%%%%%%%%%%%%%%%%%%%%%%%%%%%%%%%%%%%%%%%%%%%%%%%%%%%%%%%%%%%%%%%%%%%%%%%%%%%%%%%%% 
In what follows, we consider a thin-film superconductor, which is interfaced with a ferromagnetic layer along the $z$-axis (see Fig.~\ref{Fig1}). The magnetization $\mathbf{m}$ of the ferromagnet describes a conical spin texture with a wavevector lying in the plane of the interface (in the present work, we choose the helical wavevector to point along the $x$-axis). The exchange field of the ferromagnet is induced into the superconductor by proximity coupling.  For simplicity, we assume that the strength of the induced Zeeman field $\mathbf{h}$ is constant through the superconducting layer. The orientation of this Zeeman field is assumed to have no spatial variations along the $z$-axis, whereas the orientation in the $xy$-plane is 
collinear with the magnetization of the overlying ferromagnetic layer.

 To derive the GL free-energy functional, we start by formulating the quantum partition function $ \mathcal{Z}$ of the superconductor in terms of a functional integral over the Grassmann fields $\{ \bar{\psi}_{\sigma} (\mathbf{r}, \tau), \psi_{\sigma} (\mathbf{r}, \tau) \}$:~\cite{Altland:book} 
 \begin{equation}
 \mathcal{Z} = \int D(\bar{\psi}, \psi) \exp (- S[\bar{\psi}, \psi] ) . \label{Eq:Z}
 \end{equation}
 Here, $S$ is the action $S =  \int_{0}^{\beta}{\rm d\tau} \int_V {\rm d\mathbf{r}} ( \mathcal{S}_0 + \mathcal{S}_I  )$ (where $\beta= 1/k_B T$, and $V$ is the volume), which consists of a single-particle part and an interaction part. The single-particle part is
 \begin{eqnarray}
 \mathcal{S}_0 &=&  \bar{\psi}_{\sigma} (\mathbf{r},\tau) \left( \frac{\partial}{\partial \tau} + \mathcal{H}_{\sigma \sigma^{'}}  \right) \psi_{\sigma^{'}} (\mathbf{r},\tau) , \\
 \hat{\mathcal{H}} &=&  - \frac{\nabla^2}{2m}  - \mu_F + \mathbf{h}\cdot \hat{\boldsymbol{\sigma}} , \label{Eq:ConMod}
 \end{eqnarray}
 where $m$ is the effective mass, $\mu_F$ is the chemical potential, $\mathbf{h}$ is the Zeeman field induced in the superconductor by the helimagnet, and $\hat{\boldsymbol{\sigma}}$ is a vector of the Pauli matrices. 
 The subscript of $\psi_{\sigma}$ labels the spin state. Hereafter, the hat indicates that the operators are $2\times 2$ matrices in the spin space, and we assume a summation of repeated indices. For the interaction part, we consider a potential with an on-site attractive interaction of $g >0$ between the electrons:
  \begin{eqnarray}
 \mathcal{S}_I &=&  -  g \bar{\psi}_{\uparrow}(\mathbf{r},\tau) \bar{\psi}_{\downarrow}(\mathbf{r},\tau)\psi_{\downarrow}(\mathbf{r},\tau)\psi_{\uparrow}(\mathbf{r},\tau) . \label{Eq:S_I}
 \end{eqnarray}
 We set $\hbar=1$ in the above expressions and the following intermediate calculations, but will reintroduce it in the central results at the end of this section.  

We consider a Zeeman field with a helical Bloch structure:
 \begin{equation}
 \mathbf{h}= h_0 [ \sin (\xi ), \cos (\xi ) \sin (q_s x) , \cos (\xi ) \cos (q_s x)  ] . \label{Eq:h}
 \end{equation}
Here, $q_s$ is the helical wavevector along the $x$-axis,  and $\xi$ parameterizes the tilting of the helix out of the $yz$-plane. 
A Bloch helix is the most energetically favorable texture in helimagnets stabilized by indirect electron-induced exchange interactions because it minimizes the demagnetizing field of the texture. In addition, in chiral magnets with cubic symmetry, a Bloch structure is expected. In this case, the relativistic DMI stabilizes the helix. In contrast, a N\'{e}el structure is the most likely texture in chiral magnets, where the DMI is governed by an interfacial Rashba SOC. We discuss the case with a N\'{e}el helix at the end of this section.  When $| \xi | > 0$, Eq.~\eqref{Eq:h} describes a conical texture, which is observed in helical magnets under the application of a weak external magnetic field along the axis perpendicular to the plane in which the magnetization rotates.

To facilitate calculation of the partition function, we transform to a non-uniformly rotated frame via the unitary transformation $\phi_{\sigma} = U_{\sigma \sigma^{'} } \psi_{\sigma^{'} } $.
The unitary matrix is $\hat{U}= \hat{U}_y \hat{U}_x$, where $\hat{U}_x= \cos (q_s x / 2) - i \sin (q_s x/ 2) \hat{\sigma}_x$ and $\hat{U}_y= \cos (\xi / 2) + i \sin (\xi / 2) \hat{\sigma}_y$.
The transformation untwists the helix such that the single-particle part of the action becomes translationally invariant:
 \begin{eqnarray}
 \mathcal{S}_0 &=&  \bar{\phi}_{\sigma}(\mathbf{r},\tau)\left( \frac{\partial}{\partial \tau}   +  \mathcal{H}_{u, \sigma \sigma^{'} } \right) \phi_{\sigma^{'}}(\mathbf{r},\tau) ,  \label{Eq:S0_2a} \\
 \hat{\mathcal{H}}_u &=& - \frac{\nabla^2}{2m}  - \mu_F - \frac{i}{m} \mathbf{a}\cdot \hat{\boldsymbol{\sigma}} \frac{\partial}{\partial x} + h_0\hat{\sigma}_{z} , \label{Eq:S0_2b}
 \end{eqnarray}
whereas the form of the interaction $ \mathcal{S}_I $ is unchanged.~\cite{Comment:Invariance} 
In the rotated frame, the electrons experience a fictitious SOC, which is produced by the magnetic texture when the spins of the electrons follow the local magnetization direction.    
The effective SOC is parameterized by the vector $\mathbf{a}$, which is determined by  the helical wavevector and tilt angle of the spin helix: 
 \begin{equation}
 \mathbf{a}= \frac{q_s}{2} [ \cos (\xi ), 0 , \sin (\xi )] . 
 \end{equation}
In Eq.~\eqref{Eq:S0_2b}, we disregard a constant term $\mathbf{a}\cdot\mathbf{a}/2m$, which can be absorbed into the chemical potential.

A general element of SU(2) is parameterized by three Euler angles.~\cite{Auerbach:book} Two of these angles are physical and determine the orientation of $\mathbf{h}$, whereas
the third angle (which we label as $\chi$) corresponds to a rotation about $\mathbf{h}$. 
For $\xi$ near zero, $\hat{U}$ involves only physical rotations and $\chi\approx 0$, whereas at
$\xi  = \pm \pi / 2$ the operator $\hat{U}_x$ corresponds to a rotation by  $\chi= q_s x$. 
Thus, the above gauge transformation provides a correct physical description of the texture-induced SOC for small tilt angles $|\xi | \ll \pi / 2$ only. 
An alternative unitary transformation that rotates $\mathbf{h}$ parallel to the $z$-axis and involves only the orientation of $\mathbf{h}$ is  
$\hat{U}= \mathbf{n}\cdot \hat{\boldsymbol{\sigma}} / \| \mathbf{n}  \|$ where $\mathbf{n}= \hat{\mathbf{z}} + \mathbf{h}/h_0$. However, this rotation operator generally yields a position-dependent effective SOC, which does not leave the single-particle part of the action translationally invariant. Therefore, we restrict ourselves to small tilt angles $\xi$, which is valid for weak external magnetic fields (see Sec.~\ref{App4}). 
The full dependency of the magnetoelectric coupling on the tilt angle is numerically investigated in Sec.~\ref{Sec:NumStud}.  

Via a Hubbard-Stratonovich transformation, the interaction part of the action is decoupled by introducing the bosonic field $\Delta (\mathbf{r},\tau)$, which is proportional to the 
order parameter field $\Psi  (\mathbf{r})$ in Eq.~\eqref{Eq:Fe}.
An effective action for the bosonic field is obtained by integrating out the fermionic fields (see Sec.~\ref{App1} for details). 
Near the superconducting transition, the effective action can be approximated by performing an analytic expansion in powers of $\Delta$. 
Because we aim to investigate the Lifshitz invariant induced by the spin helix, which is proportional to $|\Delta|^2$ (see Eq.~\eqref{Eq:Fme}), we expand the effective action to the second order in the bosonic field:   
 \begin{equation}
 \mathcal{Z} = \int D(\bar{\Delta}, \Delta)  \exp \left(-  \sum_{\mathbf{q} , \nu_n } \mathcal{K} (\mathbf{q}, \nu_n)  |\Delta (\mathbf{q}, \nu_n)|^2  \right) . \label{Eq:Zeff} 
 \end{equation}
 Here, $\nu_n$ are the bosonic Matsubara frequencies, and the expression for $ \mathcal{K} (\mathbf{q}, \nu_n)$ is provided in Sec.~\ref{App1}.
 
The GL theory captures the spatial variations by including terms up to the second order in the spatial gradients of the order parameter field.
Therefore, we focus on the static limit $\nu_n \rightarrow 0$ in Eq.~\eqref{Eq:Zeff} and expand the function $\mathcal{K} $ to the second order in the momentum $\mathbf{q}$:
 \begin{equation}
\mathcal{K} (\mathbf{q}, 0) \approx \mathcal{K}^{(0)} + \mathcal{K}^{(1)}_i q_i + \mathcal{K}^{(2)}_{ij} q_i q_j  . \label{Eq:ExpK} 
 \end{equation}
 The first and last terms in this expansion lead to the conventional GL free energy in Eq.~\eqref{Eq:Fe}.
 The first term produces the free-energy contribution $a | \Psi |^2$, whereas $\mathcal{K}^{(2)}$ yields the gradient term.
 The expressions for these coefficients are well known and given by:~\cite{Altland:book}
  \begin{equation}
\mathcal{K}^{(0)}= d_F \frac{T-T_c}{T_c} ; \  \mathcal{K}^{(2)}_{ij} =\frac{\eta^2 }{4 m} \delta_{ij}; \   \eta^2 \equiv \frac{7\zeta (3) \mu_F d_F }{ 6 (\pi k_B T_c)^2}.
 \end{equation}
 Here,  $d_F$ is the density of states at the Fermi energy~\cite{Comment}  and $\zeta (x)$ is the  Euler-Riemann zeta function.

 The $\mathcal{K}^{(1)}$-term represents the interesting part of the expansion \eqref{Eq:ExpK} because it determines the texture-induced Lifshitz invariant. 
We find that only the $x$-component of the momentum density $\boldsymbol{\mathcal{P}}$ couples to the spin texture:
  \begin{equation}
\mathcal{K}^{(1)}_i= \frac{1}{8} \frac{h_0\sin (\xi)}{\mu_F} \frac{q_s \eta^2}{m} \delta_{ix} . \label{Eq:K1}
 \end{equation}
 A detailed derivation of Eq.~\eqref{Eq:K1} is provided in Sec.~\ref{App2}.
 We are not aware of any previous derivations of this term.
 Eq.~\eqref{Eq:K1} characterizes the magnetoelectric coupling between the superconducting
 condensate and the spin helix. It should be noted that the specific form of the tensor coefficient $\mathcal{K}^{(1)}_x$ depends on the type of helix. For example, a N\'{e}el helix, where the Zeeman field takes the form of  
$ \mathbf{h}\sim [ \cos (\xi ) \sin (q_s x) ,  \sin (\xi ), \cos (\xi ) \cos (q_s x)  ] $, leads to a magnetoelectric coupling term $\sim h_y \mathcal{P}_x$. 
In general,  the component of the Zeeman field collinear with $(\mathbf{h}\times \partial_x \mathbf{h})_{\xi=0}$ couples to the momentum density $\mathcal{P}_x$. 
 
 By Fourier transforming back to real space and defining the order parameter field as $\Psi (\mathbf{r}) = \eta \Delta (\mathbf{r})$, we obtain the following 
 GL free-energy density:  
 \begin{equation}
\mathcal{F} =  \frac{1}{4m} (\boldsymbol{\Pi}\Psi)^{\ast}\cdot  \boldsymbol{\Pi}\Psi + a |\Psi|^2 + \frac{b}{2} |\Psi|^4 -\frac{\kappa}{2}h_x \mathcal{P}_x + \frac{B^2}{ 8\pi} . \label{Eq:GLenergy}
\end{equation}
Here, we have included the standard fourth-order term~\cite{Altland:book} and the magnetic vector potential in the momentum operator, which contains the demagnetizing field from the magnetic texture in addition to any externally applied magnetic fields. The magnetoelectric coupling parameter is $\kappa = \hbar q_s / 8\mu_F m$, whereas $a= d_F (T-T_c)/T_c \eta^2$ and $b= 3d_F \pi^2 / 2\eta^2 k_F^3$.

The resulting GL equations of Eq.~\eqref{Eq:GLenergy} are obtained by a variational minimization of the free-energy functional with respect to $\Psi^{\ast}$ and $\mathbf{A}$:
\begin{eqnarray}
\frac{\Pi^2 \Psi  }{4m} &=& -  \alpha \Psi  - \beta |\Psi |^2 \Psi + \kappa h_x \Pi_x\Psi , \label{Eq:GL1} \\
\frac{j_{s,i}}{2e} &=&  \frac{1}{4m} \left[ \Psi^{\ast} \Pi_i\Psi +  \Psi \Pi_i^{\ast} \Psi^{\ast} \right]   -  \delta_{i x} \kappa h_x |\Psi |^2 . 
\label{Eq:GL2}
\end{eqnarray}
Here, $\mathbf{j}_s= (c/4\pi) \left(\boldsymbol{\nabla}\times \mathbf{B} \right)$ is the supercurrent density.
The variation with respect to $\Psi^{\ast}$ yields two surface terms, which give the equilibrium boundary conditions for $\Psi$:  
$s_i [  \Pi_i \Psi / 2m - \delta_{ix}\kappa h_x \Psi ]= 0$ (where $\mathbf{s}$ is the surface normal). The magnetic field $\mathbf{B}$ satisfies the usual boundary conditions of Maxwell's equations. 

Eqs.~\eqref{Eq:GLenergy}-\eqref{Eq:GL2} represent the central results of this paper and provide a phenomenological description 
of a conventional superconductor coupled to a helimagnet.   

Eqs.~\eqref{Eq:GLenergy}-\eqref{Eq:GL2} clearly show that the spin texture leads to several important effects in the superconductor-helimagnet heterostructure. 
First, in agreement with the effective Hamiltonian derived for an YSR chain in Ref.~\onlinecite{Pientka:prb2013}, a helix with a conical shape leads to a non-vanishing phase gradient $\boldsymbol{\nabla} \phi$ along the helical wavevector $\mathbf{q}_s$. This phase variation is caused by the magnetoelectric coupling term in Eq.~\eqref{Eq:GL1} and the modified boundary conditions for $\Psi$. 
Second, the spin texture yields an anomalous term in the expression for the supercurrent density \eqref{Eq:GL2}, which affects the solution of the magnetic field $\mathbf{B}$. 
Third, the texture-induced anomalous term in the supercurrent also implies the existence of a reciprocal phenomenon; namely, a supercurrent-induced torque $\boldsymbol{\tau}_{\rm me} $ on the magnetization. This torque mechanism can be used to manipulate the magnetic texture using the supercurrents and opens the door to dissipationless current-driven control of the magnetization. 
The torque originates from a spin polarization of the condensate, which is produced by the supercurrent via the texture-induced SOC. The supercurrent-induced spin polarization is proportional 
to $\partial \mathcal{F}_{\rm me}/\partial \mathbf{h}$. Via the coupling term $\mathbf{h}\cdot \hat{\boldsymbol{\sigma}}$ in Eq.~\eqref{Eq:ConMod}, this spin density leads to a reactive magnetization torque of the form $\boldsymbol{\tau}_{\rm me} \propto \mathbf{h}\times \partial \mathcal{F}_{\rm me}/\partial \mathbf{h} $. 
It should be mentioned that the texture-induced  anomalous supercurrent and the resulting torque have similarities to the magneto-Josephson effect studied in 1D systems with strong SOC,
where a change of the magnetic field direction across the junction was shown to produce a (spin) Josephson current and a torque on the external magnets.~\cite{Jiang:prb13}

The physical mechanism underlying the texture-induced Lifshitz invariant in Eq.~\eqref{Eq:GLenergy} is a shift in the Fermi surface because of the magnetic texture. 
In the rotated frame, the itinerant electrons experience an effective SOC $\sim \mathbf{a}\cdot \hat{\boldsymbol{\sigma}} p_x$. Combined with the Zeeman splitting $h_0 \hat{\sigma}_z$,  this effective SOC yields an asymmetric Fermi surface 
on which optimal Cooper pairing occurs for pairs of electrons with a finite center-of-mass momentum along $x$ (Fig.~\ref{Fig1}b). This Cooper pairing leads to a spatial modulation of the superconducting phase along $x$, which is phenomenologically captured by the Lifshitz invariant \eqref{Eq:Fme2}.     
Because the spin texture is the underlying mechanism that breaks the inversion symmetry of the superconductor,
 the Lifshitz invariant vanishes in the limit $q_s \rightarrow 0$. 

In noncentrosymmetric superconductors with SOC and a uniform Zeeman field, it is well known that the supercurrent induced by the phase variation $\Psi\sim \exp(i \mathbf{Q}\cdot \mathbf{r} )$ exactly counterbalances the anomalous part generated by the SOC.~\cite{Edelstein:JETP89} 
In Sec.~\ref{App4}, we show that the same is the case in superconductor-helimagnet heterostructures with a constant Zeeman field strength $|\mathbf{h} |$. 
Using a collective coordinate description, we find that an external magnetic field along the helix is required to tilt the helix and induce a phase variation. The supercurrent induced by this 
phase variation cancels the anomalous part of Eq.~\eqref{Eq:GL2}, leading to a zero net supercurrent density.  

In contrast, several works have shown that inhomogeneities in the Zeeman field produce a non-vanishing supercurrent density in noncentrosymmetric superconductors with SOC.~\cite{Malshukov:prb16, Balatsky:prl15}
Examples of systems where this effect becomes important are conventional superconductors with magnetic impurities (e.g., YSR chains) or islands placed on the surface. 
The current contributions from the phase variation and the anomalous part will not cancel each other in these systems. Instead, the Zeeman field generates a strong supercurrent density, which even contains signatures of the superconductor's topological state.~\cite{Balatsky:prl15} Below, we show that a similar phenomenon is observed with regard to the texture-induced magnetoelectric coupling.
By studying helical ordered YSR chains, we show that conical shaped helices generate a supercurrent density, which is consistent with the above phenomenology.

%%%%%%%%%%%%%%%%%%%%%%%%%%%%%%%%%%%%%%%%%%%%%%%%%%%%%%%%%%%%%%%%%%%%%%%%%%%%%%%%%%%%%%%%%%%%%%%%%%%%%%%%%%%%%%%% 
\section{Numerical Investigation} \label{Sec:NumStud}
%%%%%%%%%%%%%%%%%%%%%%%%%%%%%%%%%%%%%%%%%%%%%%%%%%%%%%%%%%%%%%%%%%%%%%%%%%%%%%%%%%%%%%%%%%%%%%%%%%%%%%%%%%%%%%% 
We now investigate the predicted textured-induced magnetoelectric coupling by performing a self-consistent numerical calculation of the pair potential and the supercurrent in a 2D superconductor with a spin chain placed on the surface. 
A spin spiral state is stabilized in this system by either electron-induced exchange interactions~\cite{Christensen:arXiv2016} or the relativistic DMI.~\cite{Menzel:prl12}
To demonstrate the linear relationship between $q_s$ and the magnetoelectric coupling, we allow $q_s$ to vary in the following numerical calculations. 

\subsection{Model} \label{Sec:Model}
We model the 2D superconductor using the tight-binding Hamiltonian
\begin{eqnarray}
H &=& -t \sum_{\langle \mathbf{ij} \rangle} \mathbf{c}_{\mathbf{i}}^{\dagger} \mathbf{c}_{\mathbf{j}} - \mu\sum_{\mathbf{i}} \mathbf{c}_{\mathbf{i}}^{\dagger} \mathbf{c}_{\mathbf{i}} + \sum_{\mathbf{i}}  \mathbf{c}_{\mathbf{i}}^{\dagger} \left(  \mathbf{h}_{\mathbf{i}} \cdot\hat{\boldsymbol{\sigma}} \right) \mathbf{c}_{\mathbf{i}} +\nonumber  \\
& &    \sum_{\mathbf{i}} \left( \Delta_{\mathbf{i}}  c_{\mathbf{i}\uparrow}^{\dagger} c_{\mathbf{i}\downarrow}^{\dagger} +  h.c. \right)  .    \label{Eq:H0}
\end{eqnarray}
Here, $\mathbf{c}^{\dagger}_{\mathbf{i}}  = (c_{\mathbf{i} \uparrow}^{\dagger}  \  c_{\mathbf{i} \downarrow}^{\dagger} )$, where $c_{\mathbf{i} \tau}^{\dagger} $ is a fermionic creation operator that creates a particle with spin $\tau$ at lattice site $\mathbf{i}= (x,y)$; the symbol $\langle \mathbf{ij} \rangle$ implies a summation over the nearest lattice sites;  $t$ is the spin-independent hopping energy; $\mu$ is the chemical potential;~\cite{Comment2} and  $\mathbf{h}_{\mathbf{i}}$ is the Zeeman field induced by the spin texture. Locally, the Zeeman field is collinear with the magnetization of the spin chain and is determined by a discretization of Eq.~\eqref{Eq:h}.  

$\Delta_{\mathbf{i}}= g \langle c_{\mathbf{i} \uparrow} c_{\mathbf{i} \downarrow}  \rangle$ describes the superconducting s-wave pairing. 
By inserting the Bogoliubov transformation  $c_{\mathbf{i}\tau}= \sum_n [  u_{n\tau} (\mathbf{i}) \gamma_n +  v_{n\tau}^{\ast} (\mathbf{i}) \gamma_n^{\dagger} ]$ into $\langle c_{\mathbf{i} \uparrow} c_{\mathbf{i} \downarrow}  \rangle$ and evaluating the thermal average $\langle ... \rangle$, we obtain the following self-consistency condition for $\Delta_{\mathbf{i}}$~\cite{deGennes:book} 
\begin{eqnarray}
\Delta_{\mathbf{i}} &=& -\frac{g}{2}\sum_{n\tau\tau^{'}} (i\hat{\sigma}_y)_{\tau \tau^{'}} v_{n\tau}^{\ast} (\mathbf{i}) u_{n\tau^{'}} (\mathbf{i}) \left[ 1-2f (\epsilon_n) \right] .\label{Eq:delta1}
\end{eqnarray}
Here,  $f(\epsilon)$ is the Fermi-Dirac distribution, and the operators $\gamma_n^{\dagger}$ ($\gamma_n$) are Bogoliubov quasi-particle creation (destruction) operators, which represent a complete set of energy eigenstates:  $H= E_g + \sum_n \epsilon_n \gamma_n^{\dagger}\gamma_n $ ($E_g$ is the groundstate energy).
The resulting Bogoliubov-de Gennes (BdG) equations~\cite{deGennes:book} are iteratively solved (with open boundary conditions in Eq.~\eqref{Eq:H0}) with the self-consistency condition \eqref{Eq:delta1}  
until the norm $\| \Delta \| = \sqrt{\sum_{\mathbf{i}} |\Delta_{\mathbf{i}}  |^2}$ reaches a relative error on the order of $10^{-5}$.  

The current density is calculated from the eigenfunctions of the BdG equations (see Sec.~\ref{App3}).

We consider a $41\times 29$ lattice, where the helical spin chain induces a spatially varying Zeeman field in the central region $x/a \in [11, 31]$ and $y/a= 15$ ($a$ is the discretization constant). 
The Hamiltonian \eqref{Eq:H0} is scaled with the hopping energy $t$, and the chemical potential, the pairing strength,  the Zeeman splitting, and the thermal energy $k_B T$ are set to $\mu/ t= -2.0$, $g/ t = 3.0$,  $h_0/ t= 1.5$, and $k_B T/ t= 0.001$, respectively. Without the Zeeman field, these parameter values yield a superconducting gap of $\Delta/t\sim 0.46$. 

The hopping energy in Eq.~\eqref{Eq:H0} is related to a central difference discretization of the corresponding continuum model in Eq.~\eqref{Eq:ConMod} via the relationship 
$ t= \hbar^2/2ma^2$.
With $a=2.8$~nm and $m= m_e$ ($m_e$ is the electron mass), the provided parameter values correspond to the model in Ref.~\onlinecite{Klinovaja:PRL2013} describing a Fe chain on a conventional superconductor, where the Fermi energy is $\mu_F= 10$~meV, the Fermi wavevector is $k_F = 5\times10^8$~m$^{-1}$, and the Zeeman splitting is $h_0= 7.5$~meV. 
We will use these material parameters as a typical model system of a YSR chain.   

\subsection{Results and discussion} \label{Sec:Results}
%%%%%%%%%%
\begin{figure}[t!] 
\centering 
\includegraphics[scale=1.0]{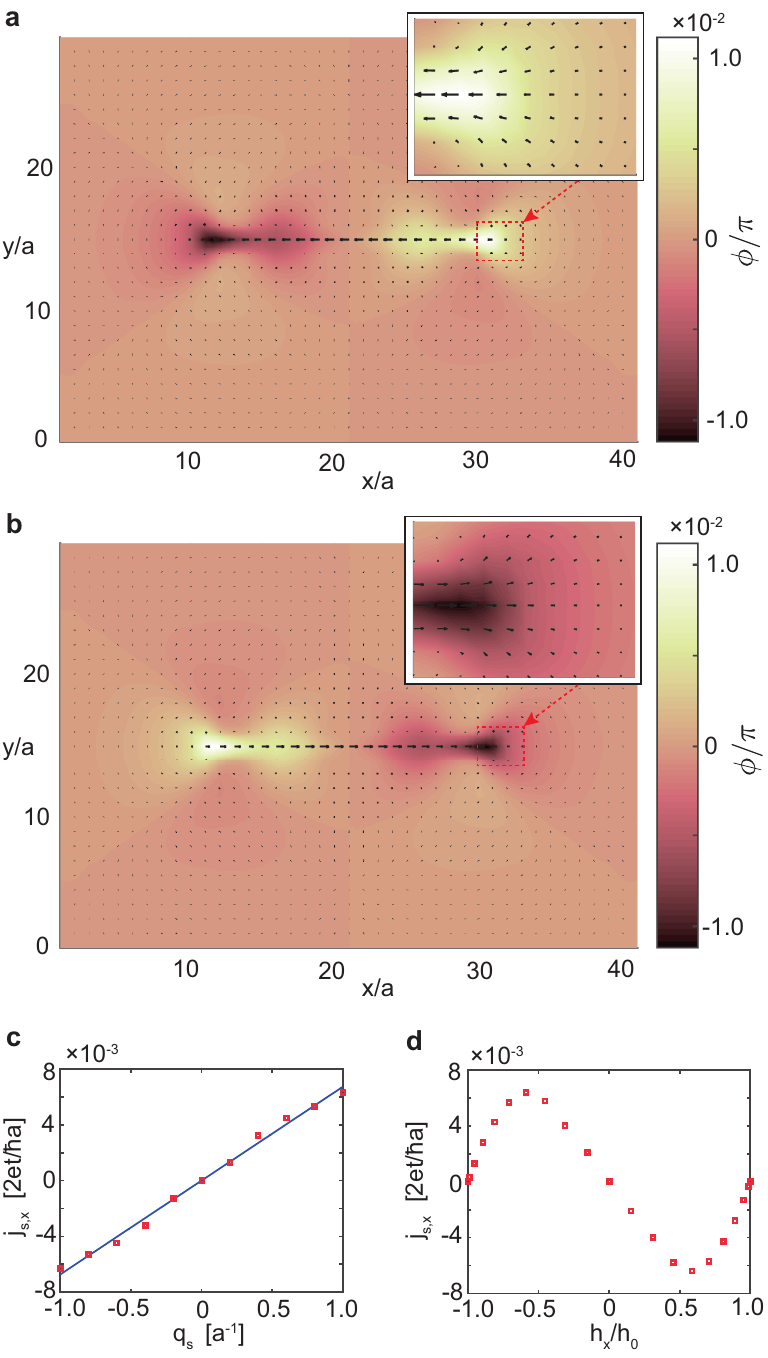}  
\caption{(Color online). (a)-(b) Superconducting phase (color) and supercurrent density (arrows) in a system that contains a helical YSR chain with $q_s a = 0.5$. 
(a) The tilt angle of the spin helix is $\xi=\pi/10$. A phase gradient and a supercurrent are induced along the YSR chain. 
(b) Switching of the tilt angle to $\xi=-\pi/10$ reverses the phase gradient and supercurrent.
The insets in (a) and (b) show the structure of the supercurrent density near the right edge of the YSR chain.
(c) The supercurrent density along the YSR chain as a function of $q_s$ (in units of $a^{-1}$) when $\xi=-\pi/10$.
The line represents the linear curve fit of the data values.
(d) The supercurrent density along the YSR chain as a function of $h_x/h_0$ when $q_s a = 0.5$.
In (c)-(d), the 2D supercurrent density is measured in units of $2et/\hbar a$.}
\label{Fig2}
\end{figure}
%%%%%%%%%%
We expect the Lifshitz invariant to produce a phase gradient and a supercurrent along the helical YSR chain that linearly depend on the 
helical wavevector and tilt angle (for small $\xi$). 
Now, we show that these predictions are confirmed by the numerical calculations.

Figs.~\ref{Fig2}a-b show the superconducting phase and the supercurrent density in the $xy$-plane for $\xi= \pm \pi/10$. As expected, a phase gradient and a supercurrent develop along the chain. The directions of these vector fields are reversed when the sign of $\xi$ is switched. 
A closer inspection of the flow pattern reveals that the current flow 
has the structure of a dipole field, in which the streamlines spread at the edges of the chain (insets in Fig.~\ref{Fig2}a-b) and circulate in a closed loop in the $xy$-plane. However, significant supercurrent density is mainly produced along the chain where the streamlines merge. 

Figs.~\ref{Fig2}c-d show the supercurrent along the chain as a function of the helical wavevector and tilt angle (parameterized by $h_x/h_0$). Clearly, a linear relationship exists between the induced supercurrent and the helical wavevector. The small deviation from the linear curve fit is because of a change of the Cooper pair density $\sim |\Delta|^2$ in the chain for different wavevectors $q_s$. 

A linear relationship is also observed in Fig.~\ref{Fig2}d for small $h_x/h_0$, which is consistent with the theory derived in Sec.~\ref{Sec:MicroDerivation}. 
Throughout the domain $ h_x/h_0 \in [-1, 1]$, the induced supercurrent has the functional form of $\sim \sin (2\xi)$.
The $ \sin (2\xi)$-dependency can be understood by considering the effective SOC  $\mathbf{a}\cdot\hat{\boldsymbol{\sigma}}p_x \sim 2i\hat{U}\partial_x \hat{U}^{\dagger} p_x $ induced by the SU(2) gauge transformation $\hat{U}= \mathbf{n}\cdot \hat{\boldsymbol{\sigma}} / \| \mathbf{n}  \|$ (here, $\mathbf{n}= \hat{\mathbf{z}} + \mathbf{h}/h_0$). In Sec.~\ref{Sec:MicroDerivation}, we show that the Lifshitz invariant originates from a shift of the Fermi surface in the rotated frame (Fig.~\ref{Fig1}b). This shift is proportional to the $z$-component of the vector $\mathbf{a}$. For the above transformation, $a_z\sim f(x)\sin (2\xi)$ where $f(x)= q_s \cos (q_s x)/ 4[1 + \cos (q_s x)\cos (\xi)]$. Clearly, $a_z$ contains a factor of $\sin (2\xi)$, which yields a vanishing effective SOC for $\xi = \pm \pi/2$ and modulates the shift of the Fermi surface (and, consequently, the magnetoelectric coupling) according to the functional form shown in Fig.~\ref{Fig2}d. 

The Lifshitz invariant makes it possible to detect the helical ordering of YSR chains via direct imaging techniques such as nanoscale scanning magnetometry.~\cite{Zeldov:NanoSQUID,Thiel:arxiv15} 
By applying a weak external magnetic field $\mathbf{B}_{\rm ext}$ along the $x$-axis, the tilting $\xi$ of the helical YSR chain can be tuned. This tilting induces a supercurrent along the chain, which above the central region of the array, produces a transverse magnetic field $\mathbf{B}_{\rm ind} || \hat{\mathbf{y}}$. 
We can estimate the magnitude of this field by considering a long wire for which the currents outside the chain are negligible, and the induced field is governed by the supercurrent $I_s= A j_s$ ($A$: cross section) along the chain. In this case, the  field measured at distance $d$ above the adatoms is $B_{\rm ind, y}= \mu_0 I_s/2\pi d$, where $\mu_0$ is the vacuum permeability. 
For the parameter values in Sec.~\ref{Sec:Model}, the electron-induced exchange interaction stabilizes a spin spiral with a helical wavevector  of  $q_s\sim  9\times 10^8$~m$^{-1}$ (adapted from Fig. 3 in Ref.~\onlinecite{Christensen:arXiv2016}).
Based on Fig.~\ref{Fig2}c, we then find that a tilt angle of $\xi=\pi /10$ produces the supercurrent density $j_s \sim 13.5$~A/m.
Assuming that the supercurrent is generated in a region of width $A\sim a$, the estimate is  $B_{\rm ind, y}\sim 0.7$~$\mu$T when $d\sim 10$~nm. 
Thus, even for a weakly tilted helical structure, we expect the helimagnet to produce a magnetic field that is detectable using imaging techniques. 
Note that the DMI is able to stabilize helices with helical wavevectors that are four to five times larger than in the above estimate.~\cite{Menzel:prl12} 
The characteristic properties of the induced field are its linear dependence on the external field for weak fields, $B_{\rm ind, y} \propto B_{\rm ext, x} $, and the limit value $B_{\rm ind, y} \rightarrow 0$ when $B_{\rm ext, x} $ approaches the critical value for ferromagnetic ordering of the spin chain.   
These considerations demonstrate that the effects of the texture-induced magnetoelectric coupling can be of particular importance in understanding self-organized topological systems and may open the door for new experimental techniques to 
probe the helical ordering of the adatom spins.  

%%%%%%%%%%%%%%%%%%%%%%%%%%%%%%%%%%%%%%%%%%%%%%%%%%%%%%%%%%%%%%%%%%%%%%%%%%%%%%% 
\section{Summary} \label{Sec:Summary}
In summary, we developed a GL theory of a conventional superconductor coupled to a helimagnet. We find that the spin texture produces a Lifshitz invariant in the GL free energy, which generates a spatial modulation of the superconducting phase in the case of a conical helix.  To investigate the effects of the texture-induced magnetoelectric coupling, we studied a helical YSR chain on a 2D superconductor and demonstrated that the Lifshitz invariant yielded a phase gradient and a strong supercurrent along the YSR chain. The induced supercurrent produces a magnetic field that is observable using imaging techniques such as nanoscale scanning magnetometry.  

We believe that the texture-induced Lifshitz invariant will be of particular importance for understanding conventional thin-film superconductors with helical ordered adatom chains on the surface. In these systems, the helical texture originates from either the interfacial SOC and the resulting DMI or the electron-induced exchange interactions. In both cases, a large helical wavevector is expected.  For helices stabilized by electron-induced exchange interactions, the helical wavevector can be comparable in magnitude to the Fermi vector,~\cite{Klinovaja:PRL2013,Vazifeh:PRL2013,Simon:PRL2013,Schecter:prb2016,Christensen:arXiv2016} whereas spin-polarized scanning tunneling microscopy measurements of Fe chains placed on Ir(001) have shown that the interfacial DMI produces an atomic scale spin spiral with $q_s\sim 4\times 10^9$ m$^{-1}$.~\cite{Menzel:prl12} The large helical wavevector results in a strong texture-induced magnetoelectric coupling, which will give rise to observable effects such as the spin-galvanic effect and induced magnetic fields.

In our study of the YSR chain, we focused on how the texture influences the equilibrium state of the superconductor. An interesting task for future studies is to investigate the back-action of the superconducting condensate on the spin system with regard to both the supercurrent-induced torque and the effect of the Lifshitz invariant on the equilibrium state of the spin texture. 
For superconductor-helimagnet heterostructures with a uniform Zeeman field strength $| \mathbf{h}|$, the analysis of Sec.~\ref{App4}  showed that an external magnetic field is required to tilt the helix.
In other words, $\xi=0$ in absence of an external magnetic field. This is not necessarily the case in systems with inhomogeneities in the Zeeman field. 
For example, unlike the system studied in Sec.~\ref{App4}, we demonstrated in Sec.~\ref{Sec:NumStud} that a supercurrent develops along a conical YSR chain. This supercurrent will, via the texture-induced SOC, produce a spin density $\sim\partial \mathcal{F}_{\rm me}/\partial \mathbf{h}$, which yields the torque $\boldsymbol{\tau}_{\rm me} \propto \mathbf{h}\times \partial \mathcal{F}_{\rm me}/\partial \mathbf{h}$ on the helix.
For the Bloch helix \eqref{Eq:h}, the supercurrent-induced spin density acts as an effective field along $x$ (according to Eq.~\eqref{Eq:Fme2}).  
An intriguing question is whether the resulting torque $\boldsymbol{\tau}_{\rm me}$ affects  the equilibrium state of the YSR chain and leads to the formation of a conical structure even in the absence of an external magnetic field. 
The action of this torque will be balanced by the anisotropy field induced by the demagnetizing field or the SOC.
Note that a conical equilibrium state requires the thickness of the ferromagnetic layer to be smaller than the transverse spin decoherence length of the ferromagnet such that the torque
$\boldsymbol{\tau}_{\rm me}$ acts on the entire helimagnetic system (not just a thin layer close to the superconductor-ferromagnet interface).
A proper treatment of this problem requires a minimization of the free energy of the total hybrid system.  
      
%%%%%%%%%%%%%%%%%%%%%%%%%%%%%%%%%%%%%%%%%%%%%%%%%%%%%%%%%%%%%%%%%%%%%%%%%%%%%%% 

%%%%%%%%%%%%%%%%%%%%%%%%%%%
\section{Acknowledgments}
%%%%%%%%%%%%%%%%%%%%%%%%%%%
K.M.D.H. would like to thank Mark S. Rudner and Michael Schecter for stimulating discussions.

\appendix

\section{Derivation of Effective Action}\label{App1}

The interaction part of the action is decoupled by introducing  the bosonic field $\Delta (\mathbf{r}, \tau)$ via the Hubbard-Stratonovich transformation~\cite{Altland:book}
\begin{widetext}
 \begin{equation}
 \exp \left( -\int {\rm d\tau}  {\rm d\mathbf{r}}\mathcal{S}_I   \right) = \int D(\bar{\Delta}, \Delta) \exp \left(-  \int {\rm d\tau}  {\rm d\mathbf{r}} \left[ \frac{1}{g}  | \Delta (\mathbf{r}, \tau) |^2  - \bar{\Delta} (\mathbf{r}, \tau) \phi_{\downarrow} (\mathbf{r}, \tau)\phi_{\uparrow} (\mathbf{r}, \tau)   - \Delta (\mathbf{r}, \tau) \bar{\phi}_{\uparrow} (\mathbf{r}, \tau) \bar{\phi}_{\downarrow} (\mathbf{r}, \tau)   \right]  \right) . \label{Eq:HS}
 \end{equation}
Here, the complex field $\Delta (\mathbf{r}, \tau)$ is proportional to the order parameter field $\Psi  (\mathbf{r})$ in Eq.~\eqref{Eq:Fe}.
Using \eqref{Eq:HS} and introducing the Nambu spinor $\boldsymbol{\phi} = [\phi_{\uparrow}, \phi_{\downarrow}, \bar{\phi}_{\downarrow},   -\bar{\phi}_{\uparrow} ]^T$, the partition function in Eq.~\eqref{Eq:Z} can be written as 
 \begin{equation}
 \mathcal{Z} = \int D(\bar{\phi}, \phi)  \int D(\bar{\Delta}, \Delta)  \exp \left(-  \int {\rm d\tau}  {\rm d\mathbf{r}}\left[  \frac{1}{g}  | \Delta (\mathbf{r}, \tau) |^2 - \frac{1}{2} \bar{\boldsymbol{\phi}} (\mathbf{r}, \tau) \left[ \boldsymbol{\mathcal{G}}^{-1} \left(\mathbf{I} + \boldsymbol{\mathcal{G}} \boldsymbol{\Sigma}  \right) \right] \boldsymbol{\phi} (\mathbf{r}, \tau)  \right]  \right) . \label{Eq:Z2}
 \end{equation}
In Eq.~\eqref{Eq:Z2}, the $4\times 4$ matrices $ \boldsymbol{\mathcal{G}}^{-1}$ and  $\boldsymbol{\Sigma}$ are given by
 \begin{equation}
 \boldsymbol{\mathcal{G}}^{-1} = 
 \begin{pmatrix}
 \hat{G}^{-1} & \hat{0} \\ 
 \hat{0} & \hat{G}^{(h) -1} 
 \end{pmatrix} , \ \ 
 \boldsymbol{\Sigma} =
 \begin{pmatrix}
 \hat{0}  & \Delta \hat{I} \\ 
 \bar{\Delta} \hat{I} & \hat{0}  
 \end{pmatrix} ,
 \end{equation} 
 where $\hat{G}$ and $\hat{G}^{(h)}$ are the non-interacting Green's functions for the particles and holes, respectively:
 $ \hat{G}^{-1} = -\partial_{\tau} -  \hat{\mathcal{H}}_u $ and  $ \hat{G}^{(h)-1} = -\partial_{\tau} +  (i\hat{\sigma}_y) \hat{\mathcal{H}}_u^{\ast}(i\hat{\sigma}_y)^{-1}$.
The operator $ (i\hat{\sigma}_y) \hat{\mathcal{H}}_u^{\ast}(i\hat{\sigma}_y)^{-1}$ denotes the time-reversal of $\hat{\mathcal{H}}_u$.  
Integrating over the fermionic fields results in an expression for the partition function in terms of the bosonic field: 
 \begin{equation}
 \mathcal{Z} = \int D(\bar{\Delta}, \Delta) \exp \left(   \frac{1}{2} {\rm Tr \ln}\ \boldsymbol{\mathcal{G}}^{-1} + \frac{1}{2} {\rm Tr \ln}\   \left(\mathbf{I} + \boldsymbol{\mathcal{G}} \boldsymbol{\Sigma}  \right)  - \frac{1}{g} \int {\rm d\tau} {\rm d\mathbf{r}} | \Delta (\mathbf{r}, \tau) |^2      \right) . \label{Eq:Z3}
 \end{equation}
 The first term in the exponent of Eq.~\eqref{Eq:Z3} represents the free energy of the normal state, and the two last terms determine the free energy of the superconducting condensate.  
 $\rm Tr [...]$ denotes the trace over the spin indices and space-time coordinates.
 Near the superconducting transition, the bosonic field is small, and the action \eqref{Eq:Z3} can be approximated by expanding the logarithm in powers of $\Delta$.  
 Our main interest is to investigate the magnetoelectric coupling induced by the magnetic texture.
 This coupling is linear in the momentum density (see Eq.~\eqref{Eq:Fme}) of the condensate and proportional to  $|\Delta|^2$. 
Therefore, we expand the action to the second order in the bosonic field, which after a Fourier transformation to the frequency and momentum representation,~\cite{Comment3} yields  the partition function 
 \begin{eqnarray}
 \mathcal{Z} &=& \int D(\bar{\Delta}, \Delta)  \exp \left(-  \sum_{\mathbf{q} , \nu_n } \mathcal{K} (\mathbf{q}, \nu_n)  |\Delta (\mathbf{q}, \nu_n)|^2  \right) , \label{Eq:Z4} \\
 \mathcal{K} (\mathbf{q}, \nu_n)  &=& \frac{1}{g} + \frac{1}{2\beta V} \sum_{\mathbf{p} , \omega_m } {\rm tr} \left[  \hat{G} (\mathbf{p} ; \omega_m )  \hat{G}^{(h)} (\mathbf{p} - \mathbf{q}; \omega_m - \nu_n) \right] .
 \label{Eq:K}
 \end{eqnarray}
 Here, $\nu_n$ ($\omega_m$) are the bosonic (fermionic) Matsubara frequencies, and the trace $\rm tr[...]$ is over the spin indices.
 \end{widetext}
 
 Eqs.~\eqref{Eq:Z4}-\eqref{Eq:K} provide the starting point for the derivation of the GL free energy to the second order in $\Psi$.
 Specifically, in the next section, we will use \eqref{Eq:Z4}-\eqref{Eq:K} to derive the Lifshitz invariant induced by the spin helix. 
 
 Note that our results do not depend on whether we perform the SU(2) transformation in Eq.~\eqref{Eq:S0_2b} before or after the integration over the fermionic fields.
 Alternatively, we could have kept the spatial dependent Zeeman field in Eq.~\eqref{Eq:ConMod}, and transformed to a rotating frame in Eq.~\eqref{Eq:Z3}.   
The effective bosonic action \eqref{Eq:Z3} is invariant under such a transformation due to the invariance of the trace $\rm Tr [...]$ under cyclic permutations:
${\rm Tr \ln}\   \left(\mathbf{I} + \boldsymbol{\mathcal{G}} \boldsymbol{\Sigma}  \right)  = {\rm Tr \ln}\  \boldsymbol{\mathcal{U}} \left(\mathbf{I} + \boldsymbol{\mathcal{G}} \boldsymbol{\Sigma}  \right)\boldsymbol{\mathcal{U}}^{\dagger}$,
where $\boldsymbol{\mathcal{U}}= [\hat{U}\ \hat{0}; \hat{0}\  \hat{U} ]$. 

 In the following discussion, it is convenient to separate $\hat{\mathcal{H}}_u$ into terms that preserve and break the time-reversal symmetry: 
 \begin{equation}
\hat{\mathcal{H}}_u = \hat{H}_u^{(0)} + h_0 \hat{\sigma}_z .
 \end{equation} 
Here, $\hat{H}_u^{(0)} = - \nabla^2/2m  - \mu_F - (i/m) \mathbf{a}\cdot \hat{\boldsymbol{\sigma}} \partial_x$ is invariant under time reversal, whereas $h_0 \hat{\sigma}_z$ switches its sign.

\section{Derivation of  Lifshitz invariant }\label{App2}  
The Lifshitz invariant \eqref{Eq:Fme} is linear in the Zeeman field and the spatial gradients of the order parameter field. 
Therefore, we start by expanding the non-interacting Green's functions to the linear order in $\mathbf{q}$ and $h_0$:  
\begin{eqnarray}
 \hat{G}(\mathbf{p}+\mathbf{q}; \omega_m)  &\approx&   \hat{G}_0(\mathbf{p}; \omega_m) + \label{Eq:G_Exp}  \\
 & &  \hat{G}_0(\mathbf{p}; \omega_m)h_0\hat{\sigma}_z \hat{G}_0(\mathbf{p}; \omega_m)   \nonumber \\
 & & + \hat{G}_0(\mathbf{p}; \omega_m)\hat{v} (\mathbf{p})\cdot \mathbf{q} \hat{G}_0(\mathbf{p}; \omega_m)  \nonumber . \\
  \hat{G}^{(h)} (\mathbf{p}+\mathbf{q}; \omega_m)  &\approx&   -\hat{G}_0(\mathbf{p}; -\omega_m) +  \label{Eq:Gh_Exp}  \\
 & & \hat{G}_0(\mathbf{p}; -\omega_m)h_0\hat{\sigma}_z \hat{G}_0(\mathbf{p}; -\omega_m) -  \nonumber \\
 & &  \hat{G}_0(\mathbf{p}; -\omega_m)\hat{v} (\mathbf{p})\cdot \mathbf{q} \hat{G}_0(\mathbf{p}; -\omega_m) \nonumber.
\end{eqnarray}
Here, $\hat{v}_i (\mathbf{p}) = \partial_{p_i}\hat{H}_u= p_i/m + \delta_{ix}\mathbf{a}\cdot\boldsymbol{\sigma}/m$ is the velocity operator, which originates from an expansion of 
the Hamiltonian to the first order in $\mathbf{q}$: $\hat{H}_u (\mathbf{p} + \mathbf{q})\approx \hat{H}_u (\mathbf{p} ) + \partial_{\mathbf{p}}\hat{H}_u (\mathbf{p})\cdot \mathbf{q} $.
For the Green's function of the holes, we used $(i\hat{\sigma}_y) \hat{\mathcal{H}}_u^{\ast}(i\hat{\sigma}_y)^{-1} =  \hat{H}_u^{(0)}  - h_0\hat{\sigma}_z$.
$\hat{G}_0$ is the Green's function determined by $\hat{H}_u^{(0)}$ and is  given by 
\begin{equation}
\hat{G}_0(\mathbf{p}; \omega_m) = \sum_{\gamma = \pm} G_0^{\gamma} (\mathbf{p};\omega_m) \hat{P}^{\gamma} , \label{Eq:G0}
\end{equation}
where $G_0^{\gamma} (\mathbf{p};\omega_m) = 1/(i\omega_m - \epsilon^{\gamma}_{\mathbf{p}})$ and $\hat{P}^{\pm}= [\hat{I} \pm \hat{\mathbf{a}}\cdot \hat{\boldsymbol{\sigma}}]/2$ ($\hat{\mathbf{a}}$ denotes the unit vector of $\mathbf{a}$) is the projection operator onto the eigenstates of $\hat{H}_u^{(0)}$ 
with the eigenvalues $\epsilon^{(\pm)}_{\mathbf{p}} = p^2/2m -\mu_F \pm q_s p \cos (\theta) /2m$. Here, and in what follows, we select the $p_x$-axis as the polar axis, and $\theta$ is the polar angle measured from this axis.

The expression in Eq.~\eqref{Eq:K}, which originates from the product of the second and third terms in the expansions \eqref{Eq:G_Exp} and \eqref{Eq:Gh_Exp}, determines the Lifshitz invariant:
\begin{equation}
\mathcal{K}^{(1)}_i = \frac{1}{2\beta V} \sum_{\mathbf{p} , \omega_m } {\rm tr} \left[  \hat{\Lambda}^{(h)} (\mathbf{p},\omega_m) \hat{\Lambda}_i^{(v)} (\mathbf{p},\omega_m) \right]. \label{Eq:K1App}
\end{equation}
Here, we have introduced the $2\times 2$ matrices
\begin{eqnarray}
 \hat{\Lambda}^{(h)} (\mathbf{p},\omega_m) &=&  \hat{G}_0(\mathbf{p}; \omega_m)h_0\hat{\sigma}_z \hat{G}_0(\mathbf{p}; \omega_m) ,  \nonumber \\ 
  \hat{\Lambda}_i^{(v)} (\mathbf{p},\omega_m)   &=&  \hat{G}_0(\mathbf{p}; -\omega_m)\hat{v}_i (\mathbf{p}) \hat{G}_0(\mathbf{p}; -\omega_m) . \nonumber 
\end{eqnarray}

Note that the magnetoelectric coupling coefficient \eqref{Eq:K1App} has the same form as in non-centrosymmetric superconductors with SOC (see Eq. (A1) in Ref.~\onlinecite{Edelstein:96}).
The difference is that $ \hat{\Lambda}_i^{(v)}$ now contains a contribution from the magnetic texture, whereas the matrix contains an anomalous velocity contribution from the SOC in non-centrosymmetric superconductors.

By applying the expression for the Green's function in Eq.~\eqref{Eq:G0}, the magnetoelectric coupling coefficient becomes 
\begin{equation}
\mathcal{K}^{(1)}_i =   \frac{h_0}{2V\beta} \sum_{\mathbf{p},\omega_m}  \mathcal{I}_{\gamma \rho}  (\mathbf{p}; \omega_m)  \chi_i^{\gamma \rho} (\mathbf{p}) , \label{Eq:K1_1}
\end{equation}      
where $\gamma ,\rho \in \{ +, - \}$ and the functions $\mathcal{I}_{\gamma \rho} $ and  $\chi_i^{\gamma \rho}$ are
\begin{eqnarray}
\mathcal{I}_{\gamma \rho}  &=&  \frac{1}{\omega_m^2 + (\epsilon^{\gamma}_{\mathbf{p}})^2 } \frac{1}{\omega_m^2 + (\epsilon^{\rho}_{\mathbf{p}})^2 } , \nonumber \\
\chi_i^{\gamma \rho} &=& {\rm tr} \left[ \hat{P}^{\gamma} \hat{\sigma}_z \hat{P}^{\rho} \hat{v}_i (\mathbf{p}) \right] .
\end{eqnarray}
The function $\mathcal{I}_{\gamma \rho} $ is rotationally symmetrical about the $p_x$-axis. With the momentum dependencies of $\chi_y^{\gamma \rho}\propto p_y$ and $\chi_z^{\gamma \rho}\propto p_z$, this symmetry implies that $\mathcal{K}^{(1)}_y=\mathcal{K}^{(1)}_z= 0$. Thus, only $\mathcal{K}^{(1)}_x$ results in a coupling between the Zeeman field and the momentum density of the condensate. 
Consequently, we expect a spatial modulation of the superconducting phase $\phi$ along $\mathbf{q}_s$. 

From the expressions for the projection operators and $\hat{v}_x$, we find that
$\chi^{+-}=\chi^{-+}= 0$ and $\chi^{++/--}= q_s \sin (\xi)/2m \pm p \cos (\theta) \sin (\xi) /m$. 
The summation over the momentum vectors  is evaluated by converting to an integral over $\mathcal{E}= p^2/2m - \mu_F$ using the relation
$(1/V) \sum_{\mathbf{p}} = (1/2)\int_{0}^{\pi} {\rm d}\theta \sin (\theta) \int_{-\mu_F}^{\infty} ({\rm d}\mathcal{E}/2\pi) (mp/\pi )$. Because the function $\mathcal{I}_{\gamma \rho}  $ strongly peaks at the Fermi energy, we can fix the factor $mp/\pi$ to its value at the Fermi energy. Integration over $\mathcal{E}$ and $\theta$ and subsequent summation over the frequencies (using $\sum_{m=0}^{\infty} 1/(2m + 1)^3 = 7\zeta (3)/8$) lead to the final result
 \begin{equation}
\mathcal{K}^{(1)}_x= \frac{7\zeta(3)}{96\pi^4} \frac{h_0 \sin (\xi) q_s k_F}{(k_B T_c)^2} .\label{Eq:K1final}
 \end{equation}   
 Eq.~\eqref{Eq:K1final} can be reformulated to the expression in Eq.~\eqref{Eq:K1} using $\mu_F= (\hbar k_F)^2 / 2m$ and $\eta^2 = 7 \zeta (3) k_F^3/24\pi^4 k_B^2 T_c^2$. 

\section{Collective Coordinate Description}\label{App4}  
In this section, we will use a collective coordinate description to analyze the equilibrium state of a superconductor-helimagnet heterostructure with a constant Zeeman field strength $|\mathbf{h}|$.
The total free energy density of the system is
 \begin{equation}
\mathcal{F}_{\rm tot} = \mathcal{F}_e + \mathcal{F}_{me}  + \mathcal{F}_m, \label{Eq:Ftot}
\end{equation}
where $\mathcal{F}_{e}$ is the free energy density of the isolated superconductor and $\mathcal{F}_{me}$ describes the magnetoelectric coupling: 
\begin{eqnarray}
\mathcal{F}_e &=&  \frac{1}{4m} (\boldsymbol{\Pi}\Psi)^{\ast}\cdot  \boldsymbol{\Pi}\Psi + a |\Psi|^2 + \frac{b}{2} |\Psi|^4  + \frac{B^2}{ 8\pi}  \\
\mathcal{F}_{me} &=& -\frac{\kappa}{2}h_x \mathcal{P}_x .
\end{eqnarray}
The spin system is assumed to be a thin magnetic film consisting of a cubic B20-type chiral ferromagnet. Examples of such magnets include MnSi, FeGe, and (Fe,Co)Si.
These magnets are commonly modeled by the free energy density~\cite{Nagaosa:nn2013}
\begin{equation}
\mathcal{F}_m = \frac{J}{2} \left(  \frac{\partial \mathbf{m}}{\partial r_i} \right)^2 - D \mathbf{m}\cdot\left( \boldsymbol{\nabla}\times\mathbf{m} \right) - M_s \mathbf{B}_{\rm ext} \cdot\mathbf{m} + U_{\rm a}. \label{Eq:Fm}
\end {equation}
Here, $\mathbf{m} (\mathbf{r})$ is a unit vector that parametrizes the local magnetization direction, $J$ is the spin stiffness describing the exchange interaction between neighboring magnetic moments, the term proportional to $D$ is the DMI, $M_s$ is the saturation magnetization, $B_{\rm ext}$ represents an external magnetic field, and $U_{\rm a}$ determines the anisotropy energy.
In what follows, we consider weak external magnetic fields, which lead to only a weak tilting of the helix, and disregard the effects of the magnetic vector potential in the free energy of the superconductor.

Our aim is to investigate how the texture-induced magnetoelectric coupling influences the equilibrium state of the superconductor and the spin helix, 
for instance, whether it can lead to the formation of a helical phase in the superconductor. Therefore, we assume that $\mathbf{m}$ and $\Psi$ take the following forms:  
\begin{eqnarray}
\mathbf{m} &=& [ \sin (\xi ), \cos (\xi ) \sin (q_s x) , \cos (\xi ) \cos (q_s x)  ] , \label{Eq:m_ColCoord}\\
\Psi  &=& \Psi_0 \exp (i Q x). \label{Eq:psi_ColCoord}
\end{eqnarray}
Here, $\Psi_0$ is a constant, whereas $\xi$, $q_s$, and $Q$ are the collective coordinates used to describe the heterostructure.
The anisotropy field is modeled by the energy density $U_{\rm a}= (K_u/2)m_x^2$, which we assume favors a Bloch helix (i.e., $K_u>0$). 
To control the tilting of the spin helix, the external magnetic field is applied along the $x$-axis: $\mathbf{B}_{\rm ext}= B_{\rm ext} \hat{\mathbf{x}}$. 
The equations for the collective coordinates are found by substituting Eqs.~\eqref{Eq:m_ColCoord}-\eqref{Eq:psi_ColCoord} into Eq.~\eqref{Eq:Ftot} and minimizing 
the resulting total free energy $F_{\rm tot}= \int \rm{d}\mathbf{r} \mathcal{F}_{\rm tot}$ with respect to the collective coordinates (i.e., $\partial_{i} F_{\rm tot}$= 0 where $i\in \{ \xi, q_s, Q  \}$):  
\begin{align*} 
0 &= [ Jq_s -D]\cos^2 (\xi) ,   \\ 
0 &=  \hbar Q - 2 m \kappa \sin(\xi) , \\
0 &= \cos (\xi) \left[  B_{\rm ext} M_s + \kappa\hbar Q \Psi_0^2 -  \mathcal{A}(q_s)\sin (\xi) \right]  .
\end{align*}
The function $\mathcal{A}$ is given by $\mathcal{A}(q_s)= K_u +  2Dq_s - Jq_s^2$.

The first of these equations yields $q_s= D/J$, which is the same as obtained by minimizing Eq.~\eqref{Eq:Fm} alone. Thus, the wavevector of the spin helix is not affected by the magnetoelectric coupling.
The last two equations determine the phase wavevector $Q$ and the angle $\xi$: 
\begin{eqnarray}
\sin (\xi ) &=& \frac{ B_{\rm ext}M_s  }{ D^2/J + K_u - 2m\kappa^2\Psi_0^2} , \\
\hbar Q &=& \frac{2 m \kappa B_{\rm ext}M_s   }{ D^2/J + K_u - 2m\kappa^2\Psi_0^2} .
\end{eqnarray}
Here, we have assumed that $| D^2/J + K_u - 2m\kappa^2\Psi_0^2 | > |B_{\rm ext}M_s |$. Otherwise,  $|\xi |= \pi/2$.
In the presence of an external magnetic field along $x$, we see that the magnetoelectric coupling leads to a phase variation of $\Psi$ along the helix, whereas
Eq.~\eqref{Eq:GL2} and the relation $\hbar Q = 2 m \kappa \sin(\xi)$ imply that the net supercurrent density is zero. 

\section{Calculation of Current Density}\label{App3}  
The supercurrent density  is derived from the Heisenberg equation ${\rm d} n_{\mathbf{i}}/ {\rm dt}= (i/\hbar) [H,n_{\mathbf{i}}]$ of the number operator $n_{\mathbf{i}}=  \mathbf{c}_{\mathbf{i}}^{\dagger} \mathbf{c}_{\mathbf{i}} $, where $H$ is the tight-binding Hamiltonian \eqref{Eq:H0}. Expressing the fermion operators $c_{\mathbf{i}\tau}$ in terms of the Bogoliubov operators using  $c_{\mathbf{i}\tau}= \sum_n [  u_{n\tau} (\mathbf{i}) \gamma_n +  v_{n\tau}^{\ast} (\mathbf{i}) \gamma_n^{\dagger} ]$ and evaluating the thermal average of the Heisenberg equation yields the following expression for the current density:
\begin{eqnarray}
 j_{s,k} (\mathbf{i})  &=& \frac{2 e t }{\hbar} \sum_{n\tau}  {\rm Im} [ u_{n \tau}^{\ast} (\mathbf{i}) D_k u_{n \tau} (\mathbf{i}) f (\epsilon_n)  + \nonumber \\ 
& & v_{n \tau} (\mathbf{i}) D_k v_{n \tau}^{\ast} (\mathbf{i}) f_h (\epsilon_n)   ]  + \nonumber \\
& &  \frac{2e}{\hbar}  \sum_{\tau\tau^{'} }   {\rm Im} \left[ \Delta_{\mathbf{i}} i\sigma_{y, \tau\tau^{'}}  \langle c_{\mathbf{i}\tau }^{\dagger} c_{\mathbf{i}\tau^{'}}^{\dagger}  \rangle  \right]. \label{Eq:NumCurrent}
\end{eqnarray}
Here, $u_{n \tau}$ and $v_{n \tau}$ are the eigenfunctions of the self-consistent solution of the BdG equations, $D_k u_{n \tau} (\mathbf{i}) = [ u_{n \tau} (\mathbf{i} + \mathbf{a}_k) - u_{n \tau} (\mathbf{i} - \mathbf{a}_k) ] /2$, where $\mathbf{a}_k$ is the lattice vector along $k\in \left\{  x,y\right\}$ and $f_h (\epsilon)= 1- f (\epsilon)$. The last term in Eq.~\eqref{Eq:NumCurrent} vanishes when the pair potential satisfies the self-consistency condition.  Otherwise, the term acts as a sink/source. 

%%% References %%%%%%%%%%%%%%%%%%%%%%%%%%%%%%%%%%%%%%%%%%%%%%%%%%%%%%%%%%%% 


\begin{thebibliography}{99}
\bibitem{Comment:Invariance}The interaction \eqref{Eq:S_I} can be written as $-(g/2) \bar{\psi}_{\alpha}(\mathbf{r},\tau) \bar{\psi}_{\beta}(\mathbf{r},\tau)\Gamma_{\alpha\beta;\lambda\mu}\psi_{\lambda}(\mathbf{r},\tau)\psi_{\mu}(\mathbf{r},\tau)$.
Here, the $\Gamma$ tensor is given by  $\Gamma_{\alpha\beta;\lambda\mu}= (1/2) g_{\alpha\beta} g_{\lambda\mu}^{\dagger}$, where $g_{\alpha\beta}$ is the $2\times2$ matrix $\hat{g}= i\hat{\sigma}_y$. 
Under a SU(2) transformation $\psi_{\alpha}= U_{\alpha\alpha^{'}}\phi_{\alpha^{'}}$, $\Gamma$ transforms as  
$\Gamma_{\alpha^{'}\beta^{'};\lambda^{'}\mu^{'}}= (1/2) (\hat{U}^{\ast T}\hat{g}\hat{U}^{\ast})_{\alpha^{'}\beta^{'}} (\hat{U}^{ T}\hat{g}^{\dagger} \hat{U})_{\lambda^{'}\mu^{'}} $.
The invariance of the $\Gamma$ tensor (and thus Eq.~\eqref{Eq:S_I}) under SU(2) follows from the relation  $\hat{U}^{ T}\hat{g}^{\dagger} \hat{U}= \hat{g}^{\dagger}$.
\bibitem{Comment} $d_F\equiv  D(0)$  where $D(\epsilon)= (1/V)\sum_{\mathbf{p}} \delta (\epsilon - \epsilon_{\mathbf{p}} )$ and $\epsilon_{\mathbf{p}}$ are the eigenvalues of
$\hat{\mathcal{H}}$ in absence of the Zeeman.
\bibitem{Comment2} The chemical potential $\mu$ in the tight-binding Hamiltonian  \eqref{Eq:H0} is related to $\mu_F$ in the corresponding continuum model \eqref{Eq:ConMod} via $\mu/t= \mu_F/t - 4$, where $ t= \hbar^2/2ma^2$ is the hopping energy obtained from a central difference discretization of the continuum model.
\bibitem{Comment3} We define the Fourier transforms of the Grassmann fields and bosonic field as 
$f ( \mathbf{r}, \tau ) = (1/\sqrt{\beta V}) \sum_{\mathbf{p},\omega_n} f ( \mathbf{p}, \omega_n ) \exp (-i\mathbf{p}\cdot\mathbf{r} - i \omega_n \tau)$  (Here, $f\in \{ \Delta, \psi_{\sigma} \}$).
\end{thebibliography}
\end{document}